\documentclass{iaus}
\usepackage{graphicx,psfig}

\title[Star Cluster Analyses] 
{Star Cluster Analyses from Multi-Band Photometry: \\
the Key Advantage of SALT's U-band Sensitivity }

\author[Fritze - v. Alvensleben {\sl et al.}]   
{Uta Fritze - v. Alvensleben, Polychronis Papaderos, Peter Anders, 
Thomas Lilly$^1$, Barbara Cunow$^2$, Jay Gallagher$^3$}%

\affiliation{$^1$Universit\"at G\"ottingen, Germany, 
$^2$Unisa Pretoria, South Africa, $^3$University of Wisconsin,
Madison, US \break email: ufritze@astro.physik.uni-goettingen.de}

\pubyear{2004}
\volume{xxx}  
\pagerange{119--126}
\date{?? and in revised form ??}
\jname{Proceedings Title IAU Symposium}
\editors{A.C. Editor, B.D. Editor \& C.E. Editor, eds.}
\begin{document}

\maketitle

\begin{abstract}
Conventionally, CMD analyses of nearby star clusters 
are based on observations in 2 passbands. 
They are plagued by considerable degeneracies between 
age, metallicity, distance (and extinction) that can 
largely be resolved by including additional passbands 
with U being most appropriate for young SCs and I or 
a NIR band for old globular clusters. For star clusters 
that cannot be resolved, integrated photometry in 
suitably selected passbands was shown to be as accurate as 
spectroscopy in independently revealing ages, 
metallicities, internal extinction, and photometric 
masses and their respective 1$\sigma$ 
uncertainties, when analysed with a dedicated 
analysis tool for their {\bf S}pectral {\bf E}nergy 
{\bf D}istributions (= {\bf SED}s) (cf. Anders 
{\sl et al.} 2004a, b, de Grijs {\sl et al.} 2003b). 
For external 
galaxies, rich star cluster populations can thus be 
efficiently analysed using deep exposures in 4 suitable 
filters. Again, the inclusion of the U-band 
significantly reduces the uncertainties in the cluster 
parameters. The age and metallicity distributions of 
star cluster systems yield valuable information about 
the formation history of their parent galaxies 
(Fritze - v. A. 2004). 
Here, we present our GALEV evolutionary synthesis models for star 
clusters of various metallicities (Anders \& Fritze 
- v. A. 2003), recently extended to include 
the time evolution of CMDs, 
the dedicated SED Analysis Tool AnalySED we developed, 
show results on the basis of HST 
data, and first results from our 
SALT PVP project on young star clusters in starburst  
and interacting galaxies.

\end{abstract}

\section{Introduction: Why Star Clusters ?}
Studies of {\bf S}tar {\bf C}lusters ({\bf SC}s) are interesting for 
many reasons: not only to learn about SC formation, evolution and 
destruction, but also because SCs are valuable benchmarks for 
stellar evolution models. 
For instance, effects and typical parameters of stellar 
rotation and binarity can well 
be studied on homogeneous (one age, one metallicity), reasonably sized 
samples of stars in clusters. Pixel-by-pixel analyses of HST ACS 
multi-band imaging data for the Tadpole galaxy, an ongoing merger, 
and other systems, with our GALEV evolutionary synthesis models have 
shown that SC formation is a major -- if not the dominant -- mode of 
{\bf S}tar {\bf F}ormation ({\bf SF}) in starbursts and galaxy mergers 
with, e.g., 70 \% of the total U-band light coming 
from young SCs in the Tadpole, not only in the main body of this 
galaxy, but all 
along its 180 kpc long tidal tail (de Grijs {\sl et al.} 2003a).

The Antennae galaxies, a beginning merger of 2 large gas-rich 
spiral galaxies similar to the Milky Way and M31 feature thousands 
of young SCs, many of them with radii and masses in the range of 
Galactic GCs (Fritze - v. A. 1998, 1999, Anders \& Fritze - v. A. 
{\sl submitted}). The HST WFPC2 images we analysed, however, only 
cover the innermost region of this nearby system at $\sim 16$ Mpc. 
SALT's large field of view will make a great difference here. 
While we can derive ages, metallicities, ${\rm E(B-V)}$, masses and 
radii individually for all the clusters with precise enough photometry 
in at least 4 passbands, we cannot say which or how many of these 
young massive compact SCs will survive long enough to be called 
{\bf G}lobular {\bf C}lusters ({\bf GC}s) in the end. 
An older version of essentially the same phenomenon, the already 
elliptical-like spiral -- spiral merger remnant NGC 7252, however, 
still features more than 150 bright and compact SCs with ages 
$\geq 600$ Myr, i.e. very probably young GCs, since they have 
already survived the critical times during the violent relaxation 
process in this system. We had predicted the possiblity that a 
secondary population of GCs could have formed in this system 
from our finding that -- as derived from the strength of the 
Balmer absorption lines in the integrated spectrum of NGC 7252 -- 
the starburst on a global scale in 
this merger had a tremendously high SF efficiency, $1-2$ orders 
of magnitude higher than in normal SF regimes and well in the range 
of SF efficiencies required for the formation of strongly bound, 
long-lived clusters by hydrodynamical cluster formation models. We  
also had predicted the metallicity for these clusters formed from 
pre-enriched gas in the spirals to be around 
${\rm (\frac{1}{2} - 1) \cdot Z_{\odot}}$ (Fritze - v. A. \& 
Gerhard 1994a, b, Fritze - v. A. \& Burkert 1995). Soon after HST 
had detected the SCs, the metallicities of two brightest ones were 
spectroscopically confirmed to be around ${\rm Z_{\odot}}$ (Whitmore 
{\sl et al.} 1993, Schweizer \& Seitzer 1993). 

This shows that SCs forming in abundance in gas-rich mergers are long-lived 
tracers of their parent galaxy's violent (star) formation history. 
 
Luminous elliptical galaxies are generally observed to show bimodal 
distributions for the optical colors (mostly V$-$I) of their GCs 
(e.g. Gebhardt 
\& Kissler - Patig 1999, Kundu \& Whitmore 2001), consistent with an 
early major merger origin of their parent galaxies but difficult to 
reconcile with hierarchical formation scenarios. While the blue peak 
is fairly universal and consistent with old and metal-poor GCs 
like the ones in the Milky Way halo, the red peak seems variable 
in position and relative height. With one color only, however, 
it is not possible to determine age and metallicity differences 
between the two populations. First optical-NIR color distributions
seem to show more structure than the optical ones 
(Kissler-Patig {\sl et al.} 2002, Puzia {\sl et al.} 2002, 
Hempel {\sl et al.} 2003). 

Using GALEV 
Evolutionary Synthesis models for SCs of various metallicities, we
showed that secondary GCs from spiral -- spiral mergers can well
explain the red ${\rm (V-I)}$ peak of the GC color distribution in
E/S0s, but also that a wide range of combinations of age and
metallicity can hide within 1 optical color peak but should split up
visibly in an optical-NIR color like ${\rm (V-K)}$ (Fritze - v.
Alvensleben 2004). GC multi-color
distributions hence provide a valuable key to the formation history
of their parent galaxies. SALT's extraordinary U-band sensitivity and
its large field of view offer a great potential here, in particular
with its future NIR arm. 

Therefore, we have settled out to study SC formation and evolution 
in a wide range of nearby galaxies with SALT, from normal galaxies 
with quiescent SF all through the strongest starbursts in massive 
gas-rich mergers. 

\section{The Role of the U-band}
Our Evolutionary Synthesis models GALEV for star clusters and 
galaxies 
calculate the time evolution of CMDs, integrated spectra,
luminosities, colors, emission and absorption line strengths from 4
Myr all through 16 Gyr (http://www.astro.physik.uni-goettingen.de/$^{\sim}$galev). 
They clearly show the great importance of the
U-band for age and metallicity determinations of young stellar
populations (cf. Fig. 1). 

\begin{figure}
\begin{center}
 \includegraphics[width=6.5cm]{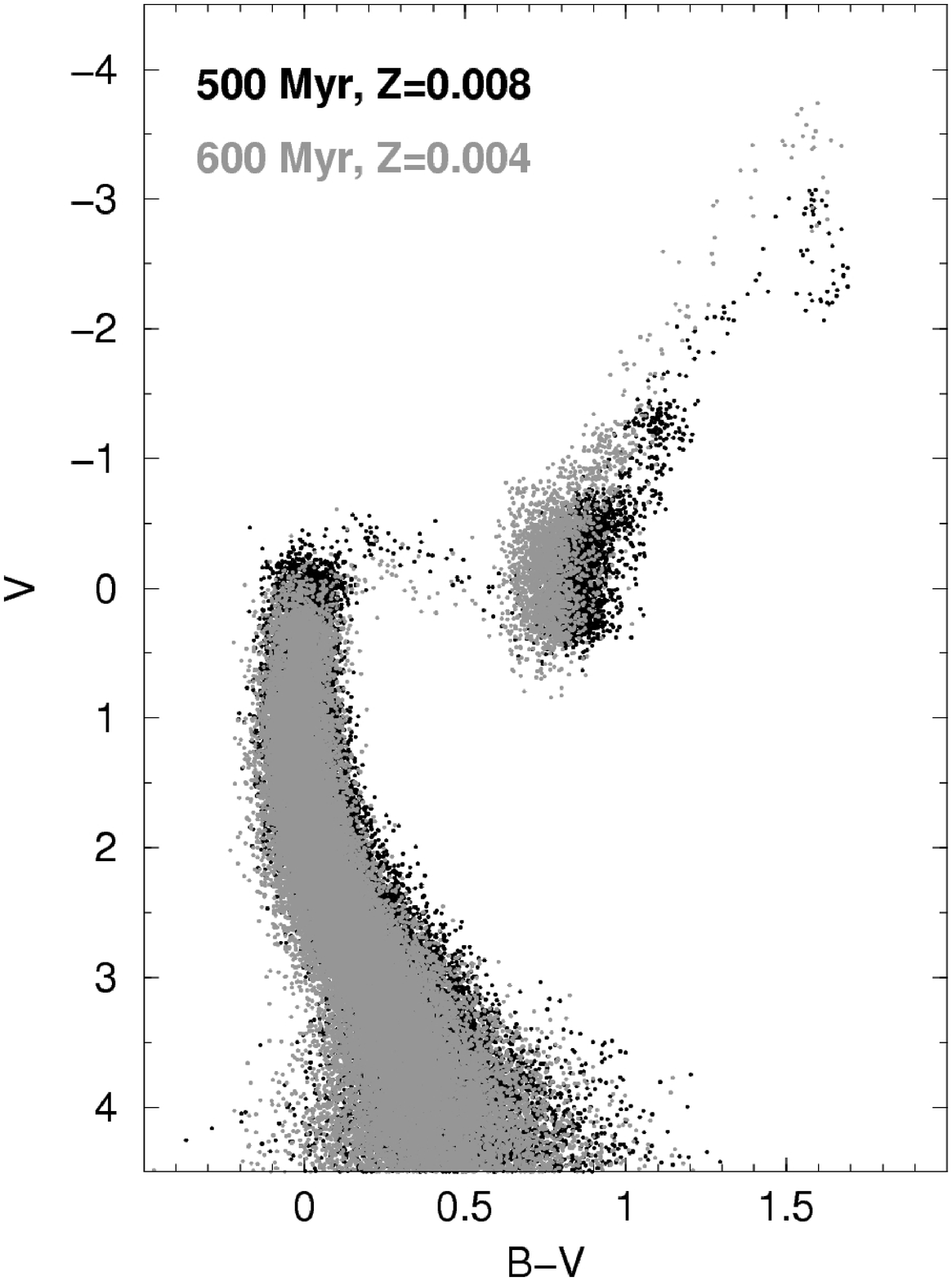}\includegraphics[width=6.5cm]{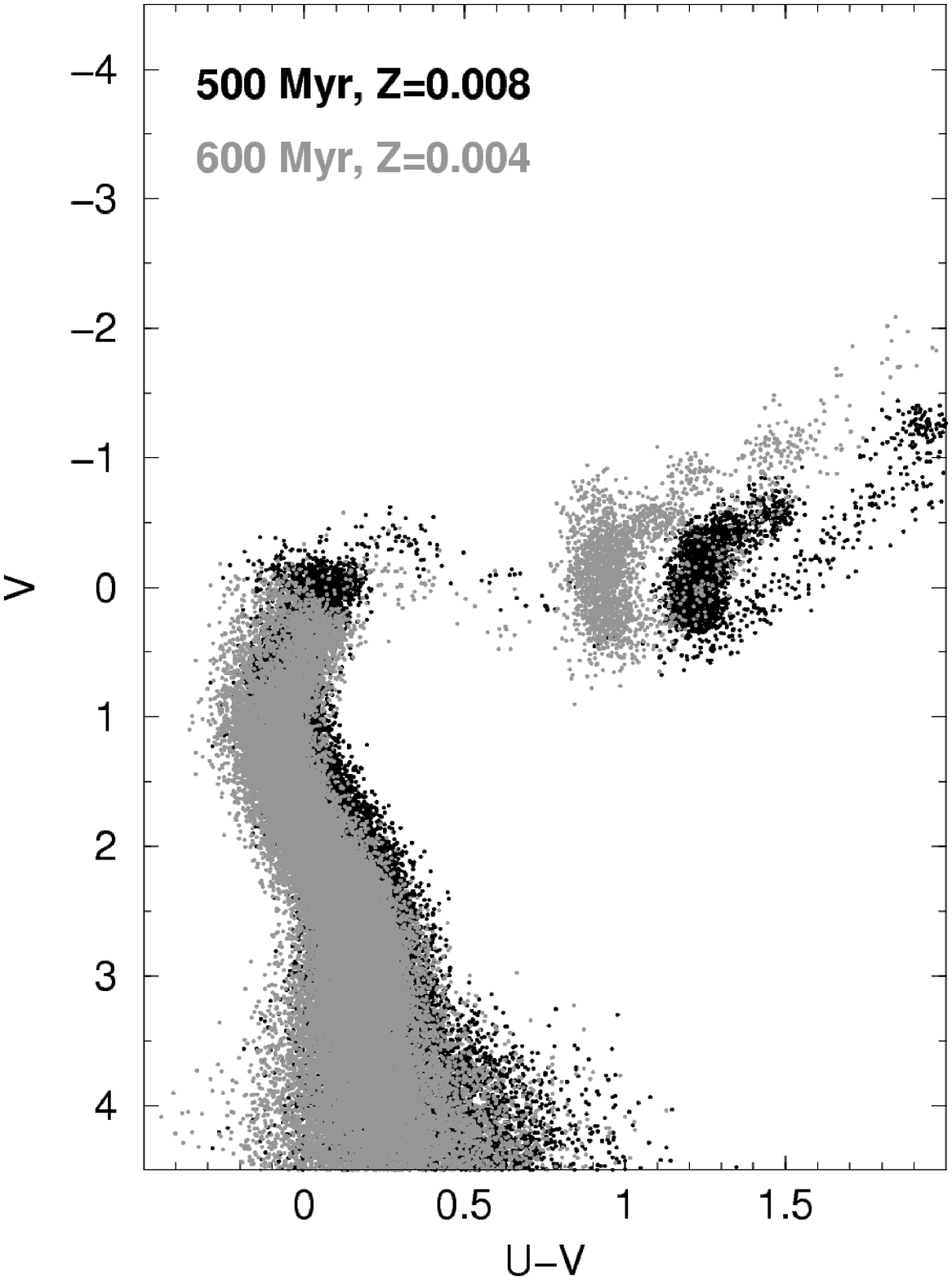}
\end{center}
  \caption{Model CMDs in V$-$(B$-$V) and V$-$(U$-$V) comparing two 
  SCs of ages 500 and 600 Myr and metallicities Z$=$0.008 and 
  0.004, respectively. While both SCs lie completely on top of 
  each other in the V$-$(B$-$V) diagram, their RGBs are very neatly
  separated in V$-$(U$-$V).}
\end{figure}

SCs are easy to model, they contain one stellar generation with one 
well-defined metallicity. For a given initial mass function of the 
stars, e.g. Salpeter, the initial mass of the cluster determines 
its initial luminosity. Soon after its formation a SC starts losing 
mass and fades due to stellar evolution in a way that depends on 
metallicity. The rate of fading, however, is different in different 
wavelength bands. Fading in all bands is fastest in early 
evolutionary stages and slows down later on. In addition to this 
stellar evolutionary mass loss and fading, real clusters also 
lose stars from the 
tail of their Maxwellian velocity distribution due to two-body
relaxation, eventually enhanced by external gravitational forces.
This mass loss and fading is not included in our modelling. At a
given age, the broad band luminosities in filters UBVRIJHK determine
the {\bf S}pectral {\bf E}nergy {\bf D}istribution ({\bf SED}) 
of a SC. We have calculated a
grid of $\sim 120 \, 000$ SEDs for SCs with metallicities in
the range ${\rm -1.7 \leq [Fe/H] \leq +0.4}$, ages in the range 4 Myr
$\dots$ 14 Gyr, and extinction values ${\rm 0 \leq E(B-V) \leq 1}$
using Calzetti {\sl et al.}'s (2000) starburst extinction law. 5
examples of SEDs for SCs of solar metallicity, ${\rm E(B-V) = 0}$, and
ages 8, 60, 200 Myr, 1 and 10 Gyr are shown in Fig. 2 The mass of a
cluster shifts the SEDs up and down. The strongest changes in the 
course of evolution are seen in the U-band relative to the longer
wavelength bands. 

\begin{figure}
\begin{center}
 \includegraphics[angle=-90,width=10cm]{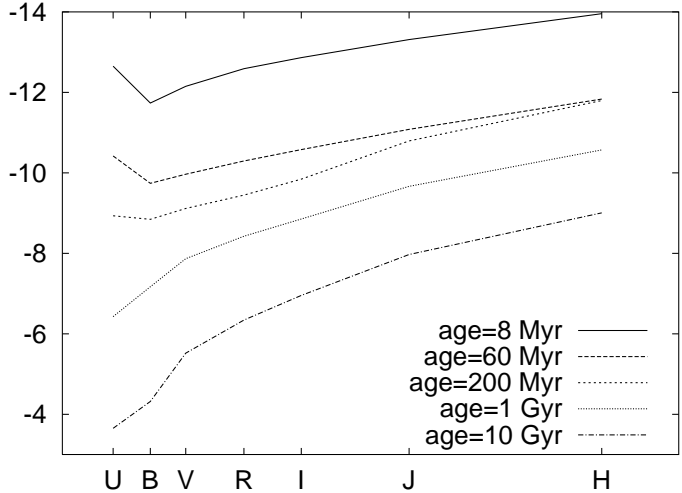}
\end{center}
  \caption{Model SEDs UBVRIH for SCs of solar metallicity, ${E(B-V) = 0}$, and
ages 8, 60, 200 Myr, 1 and 10 Gyr}.
\end{figure}

Observed SC multi-band photometric SEDs can now be compared to this grid
of model SEDs by means of a $\chi^2-$ algorithm AnalySED (Anders
{\sl et al.} 2004a) to obtain ages, metallicities, extinction values,
and masses of all individual SCs including their respective $\pm 1\sigma$
uncertainties. From these the age, metallicity and mass distributions
of SC systems can be constructed (Anders {\sl et al.} 2004b) that
then give valuable clues to their parent galaxy's SF history. 

Artificial star cluster analyses have identified the best optical
passband combinations for SC younger (older) than a few Gyr to be 
UBRI and UBVI (UBVI). In any case, the U-band is important for ages,
metallicities and extinctions of young stellar populations, while the
NIR is important for metallicities of old stellar populations. A long
wavelength basis and good photometric accuracy ($<0.1$ mag) are 
essential in any case (Anders {\sl et al.} 2004b). 

We stress that at typical photometric accuracies ($\sim 0.05$ mag),
broad band photometry with useful passband combinations is as
powerful in disentangling ages and metallicities (and extinction) as
is spectroscopy with typical S/N (cf. also Cardiel {\sl et al.}
2003). 

\section{SALTICAM PVP Observations of NGC 1487}
NGC 1487 is an ongoing merger of two spiral galaxies at a distance of
11.8 Mpc with two pronounced tidal tails. We are interested in its SC
population as described above (ages, masses, metallicities,
extinction), but also in closely examining the
merger-induced starburst over the entire system to assess the 
amount of SF that goes into cluster vs. field
star formation. 10 min U- and B-band 
exposures were taken with SALTICAM very early in the Performance
Verification Phase ($=$ PVP) with a point spread function (PSF) of 
1.96 arcsec FWHM 
and shown in Fig. 3. An overlay of B-band surface brightness contours
on the ${\rm (U-B)}$ color map clearly reveals that the
blue color peaks coincide with the highest surface brightness
regions. A flux-conserving unsharp masking technique, developed by
Papaderos (1998) applied to the U- and B-band images reveals
more than 100 compact sources fainter than 25 mag in B, a very
encouraging result at this early stage and with only part of the
total observing time we requested. An improvement in the PSF by 
about a factor 2 is expected by the time SALT/SALTICAM 
starts regular operations. 
Comparison with HST WFPC2 BVI data
on the PC chip shows that with unsharp masking SALT 
clearly resolves a wealth of SFing complexes.  

\begin{figure}
 \includegraphics[width=13.6cm,clip=]{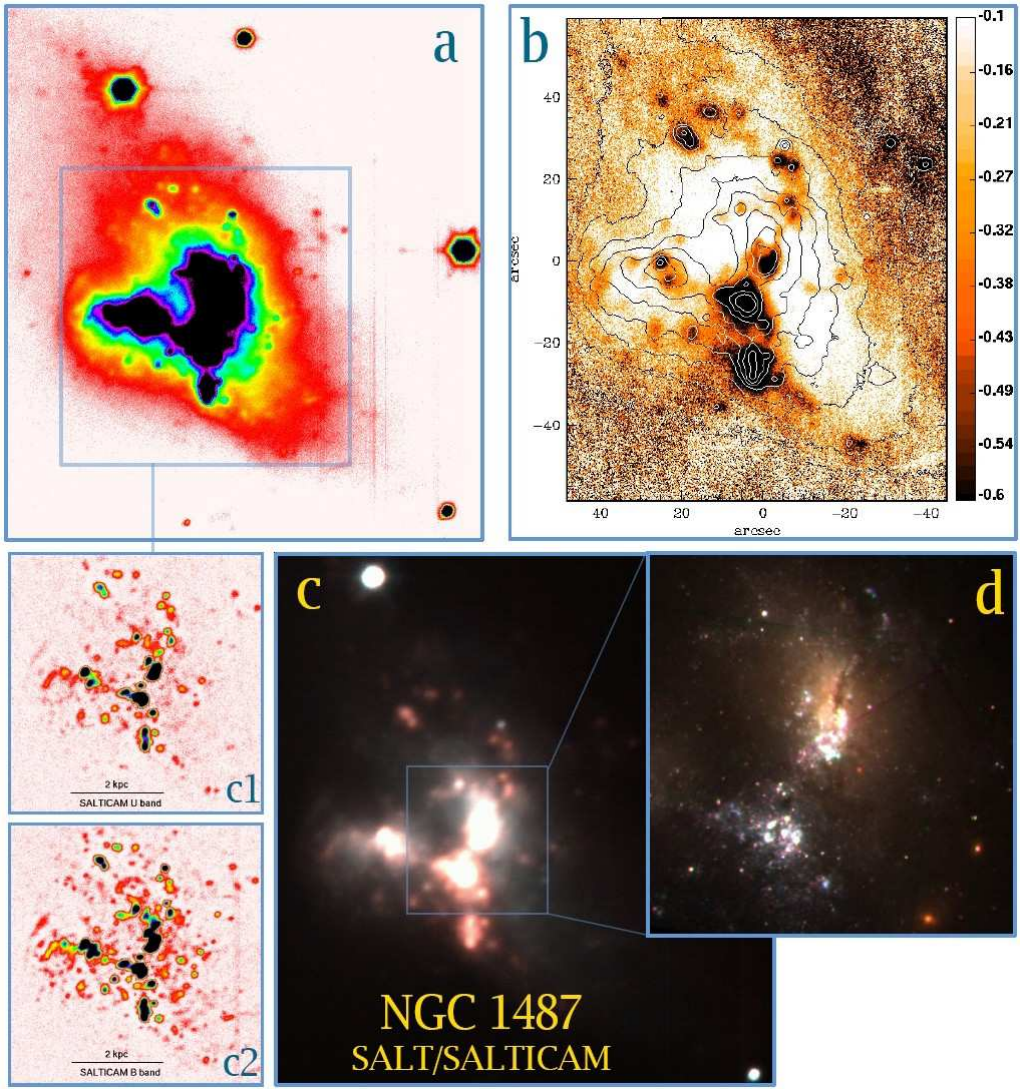}
  \caption{{\bf (a)} SALTICAM B-band image of the inner region of NGC 1487, 
{\bf (b)} B-band surface brightness contours between 20 and 25 
mag arcsec$^{-2}$ in steps of 0.5 overlaid with a (U--B) color map, {\bf (c)} {\sl true color image} of the central part of NGC 1487, 
produced by combining U and B data and {\sl unsharp-masked} versions of the respective broad band images
{\bf (c1\&c2)}, {\bf (d)} morphology of the central SFing component of NGC 1487 
as revealed by the combination of HST WFPC2 F450W, F606W, F814W archival images.}
\end{figure}

The full analysis in the way described above requires additional
observations in V and I and longer exposure times in U and B to reach
fainter limits. Once the photometry is performed, which for this PVP
object can be tied to the available HST photometry for absolute
calibration, our SED analysis tool AnalySED will immediately return 
ages,
metallicities, extinction, and masses for all the individual SCs and
star-forming complexes. 

\section{Conclusions and Outlook}
SALT's large field of view and its unique U-band sensitivity make
SALTICAM an ideal instrument for multi-band photometric analyses of
young and old stellar systems, not only star clusters as shown here,
but also for galaxies (see Fritze - v. Alvensleben {\sl et al., this
volume}). Accurate multi-band photometry allows to derive star
cluster ages, metallicities, extinctions and masses including their
respective $\pm 1 \sigma$ uncertainties (as well as galaxy types,
redshifts, star formation histories, masses and metallicities) with
accuracies comparable to those achieved in spectroscopic studies, but
reaching out to much larger distances, and provide a valuable key to
empirically constrain galaxy formation histories. 

SALTICAM has 
demonstrated its high performance already at this early stage and 
provided its first and unique data set for galaxy evolution studies.

\begin{acknowledgments}
We greatfully acknowledge the SALT astronomers and SALTICAM 
instrument team for their very efficient and successful work and for
their help with our initial data reduction problems. We also greatfully
acknowledge travel support from the DFG under Fr 916/15-1 (UFvA) and PA 1228/4-1
(PP), without which we could not have attended this conference. \\
UFvA
cordially thanks Jay Gallagher for providing her the invitation to the
inauguration of SALT. 
\end{acknowledgments}

\end{document}